\documentclass[prl,twocolumn,amsmath,amssymb]{revtex4-1}

\usepackage{graphicx}
\usepackage{bm}
\usepackage{amsmath}
\usepackage{hyperref}

\begin{document}


\title{Plasma high harmonics generation from ultra-intense laser pulses}

\author{Suo Tang}
\author{Naveen Kumar}
\email{kumar@mpi-hd.mpg.de}
\author{Christoph H.~Keitel}
\affiliation{Max-Planck-Institut f\"ur Kernphysik, Saupfercheckweg 1, 69117 Heidelberg, Germany}%


\begin{abstract}
Plasma high harmonics generation from an extremely intense short-pulse laser is explored by including the effects of ion motion, electron-ion collisions and radiation reaction force in the plasma dynamics. The laser radiation pressure induces plasma ion motion through the hole-boring effect resulting into the frequency shifting and widening of the harmonic spectra. Classical radiation reaction force slightly mitigates the frequency broadening caused by the ion motion. Based on the results and physical considerations, parameter maps highlighting optimum regions for generating a single intense attosecond pulse and coherent XUV radiations are presented.

\begin{description}
\item[PACS numbers]
52.38.-r, 42.65.Ky, 52.65.Rr,
\end{description}
\end{abstract}

\pacs{Valid PACS appear here}
\maketitle

One can not overstate the need for a powerful and coherent source of radiation operating in the extreme ultraviolet (XUV) region  due to the multitudes of its applications ranging from the novel field of attosecond physics to probing the hot and warm dense matter~\cite{Pfeifer:2006aa,Materlick:2001aa}. High harmonic generation (HHG) from linearly polarized femtosecond (fs) laser pulse interaction with solid density targets is most suitable to generate attosecond pulses and extend the coherent radiation sources into the XUV region~\cite{Dromey:2006aa,Tsakiris:2006aa,*Heissler:2012aa,Behmke:2011aa,*Braenzel:2013aa,Rodel:2012aa,*Dollar:2013aa,Gibbon:1996aa,*Teubner:2009aa}.

In general, different physical mechanisms are responsible for HHG generation, and their onset depends on the plasma density of the target, laser intensity, incident angle and plasma density gradient~\cite{Thaury:2007aa,Teubner:2009aa,Rodel:2012aa,*Dollar:2013aa}. In non-relativistic or mildly relativistic regime, $(a_0<1,$ with $a_0=eE/m_{e}c\,\omega_{l}$, $e$ and $m_{e}$ denote the charge and rest-mass of the electron respectively, $E$ and $\omega_{l}$ are the electric field and frequency of laser respectively, $c$ is the light speed in vacuum), HHG occurs due to wake created by the Brunel electrons \cite{Brunel:1987aa} inside the solid target. This mechanism is known as the coherent wake emission (CWE)~\cite{Quere:2006aa} and in this case the harmonic spectrum is limited by the local plasma frequency $\omega_{p}=(4\pi e^{2}n_{e}/m_{e})^{1/2}$ where $n_{e}$ is the local plasma electron density. In relativistic regime ($a_0 \ge 1$), HHG can be explained with the relativistic oscillating mirror (ROM) ~\cite{Bulanov:2013aa,Lichters:1996ab,Baeva:2006aa}, coherent synchrotron emission from electron nanobunches ~\cite{Brugge:2010aa,*Dromey:2012aa,*Dromey:2013aa} and relativistic electron spring models ~\cite{Gonoskov:2011aa}. In the  ROM model, the oscillating overdense layer of plasma electrons reflects the incoming laser pulse. The phase of this reflected laser pulse, owing to nonlinear interaction with the plasma,  is strongly  modulated and contains higher harmonics \cite{Bulanov:2013aa,Lichters:1996ab}.  It maybe noted that one can also produce x-ray radiation from the laser interaction with gas jet targets and underdense plasmas by the frequency up-conversion \cite{Popmintchev1287,*Nomura:2009aa,*Chang:1997aa}  and betatron radiation mechanisms \cite{Corde:2013aa}, respectively.

In comparison to gas harmonics, the underlying advantage of the fs laser-solid interaction for HHG is to be able to use higher incident laser intensity. Currently, laser systems with intensities $I_l\ge 10^{21}$W/cm$^2$ are either available or are on the horizon~\cite{Yanovsky:08,*eli:kx}. In this ultra-relativistic regime of HHG, the plasma ion motion and radiation reaction (RR) force~\cite{Landau:2005fr,Di-Piazza:2012uq,*Sokolov:2009aa,*Tamburini:2010aa,*Chen:2011aa,*tamburiniNIMA11,*Kumar:2013aa,*Vranic:2014aa,*Wallin:2015aa,Capdessus:2012aa} must instructively be taken into account. The plasma ion motion originates from the hole-boring (HB) effect~\cite{Schlegel:2009aa,*Robinson:2009aa,*Wilks:1992aa}, in which the laser pressure pushes the plasma target surface inwards creating a double-layer structure. The electron layer in this structure oscillates around the ion layer emitting high-harmonics but the structure itself has a slow motion inside the target. The HB effect also leads to strong compression of the plasma density at the laser-target interface necessitating the inclusion of electron-ion collisions  in plasma dynamics. While the plasma ion motion and collisions affect the electron layer dynamics, radiation reaction force changes the laser energy partition among electrons, ions and radiation in the plasma \cite{Capdessus:2012aa}. 

This Letter studies, for the first time, HHG by including the effects of plasma ion motion, electron-ion collisions and radiation reaction force together in the ultra-relativistic laser-plasma interaction dynamics. The Doppler shift of the incident laser frequency, arising due to the HB effect, in the rest frame of the target leads to non-integer harmonics being generated in a laboratory frame of reference~\cite{Thaury:2010aa,*Welch:2015aa}. We show that this inevitable frequency shift leads to the widening of the harmonic peaks resulting in a large frequency bandwidth in the harmonic spectra. We analyze analytically the widening of the harmonic spectra and validate it by particle-in-cell (PIC) simulations. We then discuss the effect of RR force on HHG by employing the Landau-Lifshitz prescription  of RR force \cite{Landau:2005fr} in PIC simulations. Based on these results and physical considerations, parameter maps  relating the laser intensity with the plasma density are generated, which highlight the regions suitable for both a single intense attosecond pulse and coherent XUV radiation generations for different plasma targets and incidence angles.

Essentially the physical mechanism behind the frequency broadening is the dynamic HB effect. During the initial stage of interaction, the HB velocity is not constant.This causes variable frequency shifts and since the whole spectrum is a superposition of variable frequency shifts, it results into the broadening of the harmonic spectra.  To estimate the broadening, we proceed by recalling the peak frequency shift in the laboratory frame of reference~\cite{Schlegel:2009aa,*Robinson:2009aa,*Wilks:1992aa,*Thaury:2010aa,*Welch:2015aa} which reads as
\begin{equation}
\delta\omega_{n}=n\omega_{l}-\omega_{n}^{'}=n\omega_{l}\frac{2\beta_{is}}{1+\beta_{is}}.
\label{eq1}
\end{equation}
where $\omega_l$ is the incident laser frequency in the laboratory frame of reference. The HB velocity $\beta_{is}$ is defined as $\beta_{{is}}=B/(1+B)(1-X_{h})$, $B=(1+m_{i}/Zm_{e})^{-1/2}(I\cos^{4}\theta/n_{e})^{1/2}$, $I$ is the incident laser intensity, and $X_{h}=n_{h}(2\gamma_{h}-1)/(4I\cos^{2}\theta)$ denotes the rate of energy absorbed by the hot electrons~\cite{Levy:2013aa,*Levy:2014aa}. The average hot electrons energy and  density can be approximated as $\gamma_{h}=<a_{s}^{2}>/2$, $a_{s}=2a_{0}\cos^{-1}\theta/\sqrt{n_{e}}$ is at the surface~\cite{Lichters:1996ab}, and $n_{h}=n_{rc}/3$ respectively. The relativistic critical density being $n_{rc}=\cos^{2}\theta\sqrt{1+<p^{2}>}=\cos^{2}\theta\sqrt{1+I}$, gives $X_{h}\approx \sqrt{I}/(3 n_{e} \cos^{2}\theta)$. From now onwards, we use the dimensionless quantities: $n_e=n_e/n_{c}$, $t=\omega_{l}t$, $x=k_{l}x$, $\beta=v/c$, $\omega_n=\omega_n/\omega_{l}$, $I_n=I_n/I_{r}$, where $n_c = 1.742\times 10^{21}\text{cm}^{-3}$,\,$I_{r}=c(m_{e}c\omega_{l}/e)^{2}/4\pi=4.
276\times10^{18}\,\textrm{W/cm}^{2}$.
\begin{figure}
  \includegraphics[width=0.50\textwidth,height=0.45\textwidth]{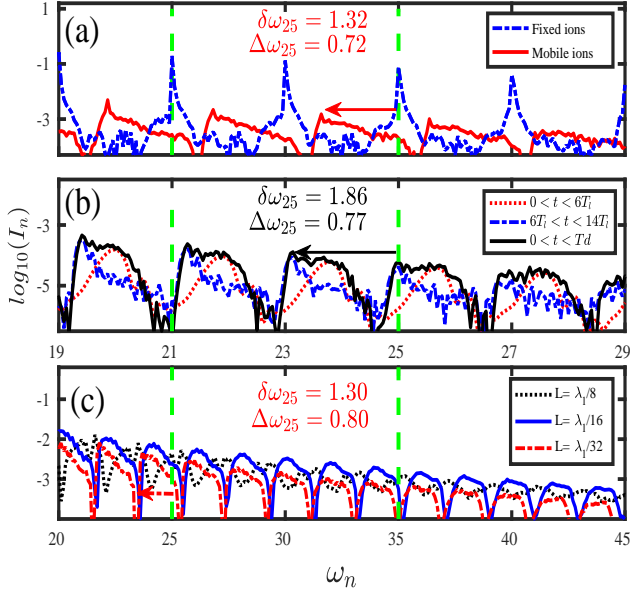}
  \caption{(color online) 1D PIC simulations of HHG. The $x$ and $y$ axes represent harmonic numbers and intensity $I_{n}$, respectively. The frequency shift (deviation from the vertical green dashed lines) is clearly seen. (a) With and without ion motion at plasma density $n_{e}=200$ and with a constant laser amplitude $a(t)=a_0$ for $0<t<T_d$. (b) With a laser pulse of temporal profile $a(t)=a_{0}\sin^{2}(\pi t/T_{d})$, where $T_{d}=20T_{l}$, at plasma density $n_{e}=80$. (c) HHG with different plasma density gradients $L$. The maximal plasma density is $n_{e}=200$ with the same laser profile as in (a) except $T_{d}=5T_{l}$. The vertical green dashed lines correspond to integer harmonics from selection rules defined in \cite{Lichters:1996ab}. }\label{Fig.1.}
\end{figure}
\begin{figure}
\centering
  \includegraphics[width=0.50\textwidth,height=0.33\textwidth]{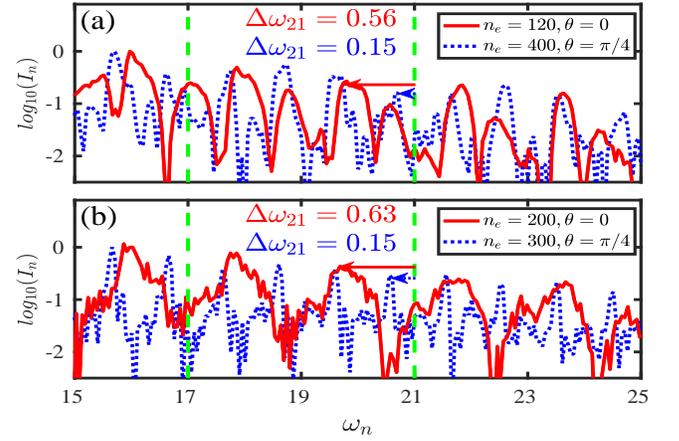}
  \caption{(color online) 2D PIC simulations with different targets and incident angles. The intensities are normalized to the intensity of the $17$th harmonic for normal incidence and the $16$th harmonic for oblique incidence. (a) Gold plasma with density gradient $L=\lambda_{l}/8$. (b) Carbon plasma with $L=\lambda_{l}/16$. The laser pulse has temporal $a(t)=a_{0}(\tanh((t-T_s)/W)-\tanh((t-T_e)/W))/2$ and transverse $a(t,y)=a(t)\exp(-y^{2}/\sigma^{2})$ profiles, where $a_{0}=40$, $W=T_{l}=\lambda_{l}/c$ and $\sigma=4\lambda_{l}$. The laser pulse has maximum intensity from $T_s=5T_{l}$ to $T_e=13T_{l}$.}\label{Fig.2.}
\end{figure}

We wish to emphasize that this peak frequency shift is indeed important for both lower and higher order harmonics. While the large HB velocity is bound to cause substantial frequency shifts, even the small HB velocity induces inevitably large frequency shifts in higher-order harmonics. Since the HB effect is not a stationary effect, this frequency shift can lead to broadening of the harmonic spectra. Considering that the maximum frequency shift and the highest intensity of the harmonics occur at the highest intensity of the laser pulse, one can scale the width (FWHM) of each harmonic quantitatively as half of the peak frequency shift $\Delta\omega_{n}=\delta\omega_{n}/2$. Fig.\ref{Fig.1.} shows 1D PIC simulation results of HHG performed with the EPOCH PIC code ~\cite{Arber:2015aa} in which a linearly polarized laser pulse, with wavelength $\lambda_{l}=0.8\mu m$ is normally incident on a preionized plasma target. Fig.\ref{Fig.1.} (a) shows the frequency shift $(\delta\omega_n)$ and  harmonic spectrum broadening $(\Delta\omega_n)$ with and without the ion motion for a constant laser pulse amplitide.  For the parameters in Fig.\ref{Fig.1.} (a) $I=800$ one get $\delta\omega_{25}=1.30$, $\Delta\omega_{25}=0.65$, while for Fig.\ref{Fig.1.} (b) peak laser intensity $I=800$ and average field $<a>=20$, one obtains $\delta\omega_{25}=1.95$, $\Delta\omega_{25}=0.97$ from Eq.\eqref{eq1}. These values match very well with the PIC simulation results. The slight discrepancy between analytical estimates and PIC simulation results can be attributed to the rough analytical estimation of the hot electron generation. Figs.\ref{Fig.1.} (b) and (c) show shift and broadening in the case of a laser pulse with temporal profile and plasma density gradient respectively including the ion motion. Apart from HB induced frequency broadening, two other effects are also responsible for broadening the harmonic spectrum: first, because of the laser temporal profile also~\cite{Behmke:2011aa,*Braenzel:2013aa}, harmonics generated by different parts of the laser pulse can have different frequency shifts, leading to frequency broadening  as shown by the solid black line in Fig.\ref{Fig.1.}(b). Second, harmonic spectra can also be broadened due to variable frequency shifts arising from the plasma density gradient \emph{i.e.} $n_e=n_0\,\textrm{exp}(x/L)$. Fig.\ref{Fig.1.}(c) depicts this broadening and one can clearly see that steep density gradient results into a narrower spectrum ( frequency shift and width for density gradient $L=\lambda_l/32$ case (red dash-dotted line) are same as in Fig.~\ref{Fig.1.}(a) which has a step-function density profile) while longer density gradient leads to widening of the harmonic spectrum.  One may note that the ion motion induced broadening dominates over the last two mechanisms and it occurs in the first few cycles of the laser pulse and therefore can not be mitigated by resorting to few cycle laser pulses for HHG as shown by the red dotted line in  Fig.\ref{Fig.1.}(b).  Finally, Fig.\ref{Fig.2.} shows 2D PIC simulation results of HHG on fully ionized gold (Au, $A/Z=197/79$) and carbon (C, $A/Z=12/6$) plasma targets with plasma density gradient. For the (a) gold and (b) carbon plasma targets, Eq.\eqref{eq1} yields the frequency bandwidth of the $21$st harmonic as (a)
 $\Delta\omega_{21}=0.72\omega_{l}, (\theta=0)$,  $\Delta\omega_{21}=0.20\omega_{l}, (\theta=45^{\circ})$ and (b) $\Delta\omega_{21}=0.61\omega_{l}, (\theta=0)$,  $\Delta\omega_{21}=0.25\omega_{l}, (\theta=45^{\circ})$ respectively and they match well with PIC simulation results in Fig.\ref{Fig.2.}.  Here we only show harmonics up to order $n\le 25$ since the peak frequency shift and  widening are expected to be large for higher-order harmonics.

\begin{figure}
  \includegraphics[width=0.48\textwidth,height=0.35\textwidth]{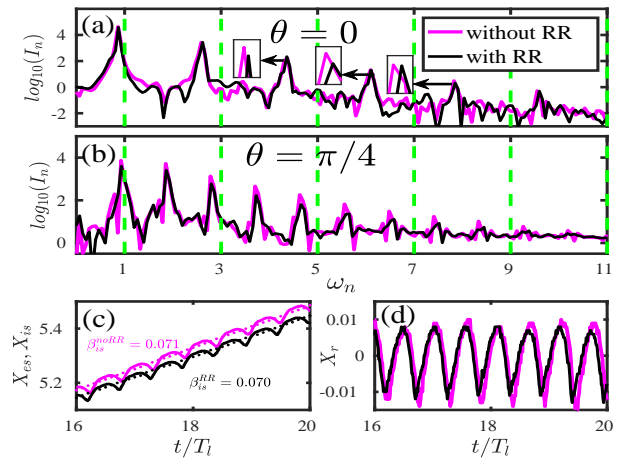}
  \caption{(color online) 2D PIC simulations of HHG. The black (dark) lines accounts for RR force while the magenta (light-grey) lines do not. (a) normal incidence $\theta=0$; (b) oblique incidence $\theta=\pi/4$. (c) Transverse motion ($y$-axis) of the electron ($X_{es}/\lambda_l$, solid lines) and ion ($X_{is}/\lambda_l$, dotted lines) layers. The electron and ion layers are defined at the position with density $n_{e}=Zn_{i}=a_{0}n_{c}$. (d) Relative position ($y$-axis) of the electron layer, $(X_{es}-X_{is})/\lambda_l$, with respect to the ion layer. (c), (d) are for $1$D normal incidence case. The laser has same profile as in Fig.\ref{Fig.1.}(a) except $a_{0}=250$ and plasma density is $n_{e}=1100n_{c}$.} \label{Fig.3.}
\end{figure}

At higher laser intensity $I_l\gg10^{22}$W$/$cm$^2$, the effect of RR force~\cite{Landau:2005fr,Di-Piazza:2012uq} is important as it influences the electron motion, partitioning of the laser energy among different particles \cite{Capdessus:2012aa}, and consequently the harmonic generation. Figs.~\ref{Fig.3.}(a) and (b) show 2D PIC simulation results depicting the influence of RR force on HHG. One can see that the peak frequency shift and the frequency widening of the harmonics is smaller with the RR force effect. This difference, albeit smaller, is clearly noticeable (see inset of Fig.~\ref{Fig.3.}(a)). Essentially RR force leads to redistribution of the laser energy among different species of particles (electrons, ions and photons) which enhances the laser energy absorption and accordingly decreases the HB velocity~\cite{Capdessus:2012aa,Schlegel:2009aa,*Robinson:2009aa} as shown in Fig.~\ref{Fig.3.}(c). However, the intensity of the harmonics emitted is also reduced by RR force in an oblique incidence case. This can be understood in the following way: HHG depends strongly on the backward motion of the electron layer towards the laser pulse. For the backward motion of the electrons, RR force is stronger and it tends to slow down the electron layer movement as shown clearly in Fig.~\ref{Fig.3.}(d). In an oblique incidence case, the backward motion can be accelerated by the $E_{x}$ component of the laser field thereby enhancing the RR force effects. Consequently the intensity of the generated harmonics can be slightly reduced in the oblique incidence case. Hence, the RR effects can slightly reduce the harmonic intensity, but at the same time slightly improve the frequency bandwidth of the generated harmonic.

This widening of the harmonics spectra caused the HB induced frequency shifts can produce a quasi-continuous frequency spectrum, a prerequisite for generating an intense isolated attosecond pulse~\cite{Tsakiris:2006aa,*Heissler:2012aa}. Though this frequency broadening can limit the temporal coherence of the high-frequency XUV radiation. This juxtaposition can be exploited to create a parameter map (laser intensity vs plasma density)  where different regions of the parameter map correspond to different applications \emph{e.g.} coherent XUV radiation and single attosecond pulse generations. The line that separates the two regions corresponds to a case when the frequency broadening equals the laser frequency. This implies $\Delta\omega_n^{\text{max}}/\omega_l=d$, where $d=1, 2$ for oblique ($p$-polarization) and normal incidences respectively, on taking into account the selection rules for HHG~\cite{Lichters:1996ab}. On using Eq.(\ref{eq1}), one gets $\beta^{\text{max}}_{{is}}=d/(n-d)$, 
yielding a bound on plasma density as
\begin{equation}
n^{\text{min}}_{e1}=\frac{(n-2d)^{2}I\cos^{4}\theta}{d^{2}\left(1+{m_{i}}/{Zm_{e}}\right)}\left(1-\frac{n-d}{n-2d}X_{h}\right)^2.
\label{eq2}
\end{equation}
For densities lower than Eq.\eqref{eq2}, harmonic orders higher than $n$ overlap with each other producing a quasi-continuos frequency spectrum while at higher densities, one gets sharp harmonic peaks, which can be spectrally filtered to produce a high-frequency radiation source with high temporal coherence. Here and below we do not taken into account RR force as it does not strongly affect the frequency shift of high-harmonics as shown in Fig.~\ref{Fig.3.}. 

The intensity of the generated harmonics is also crucial for intense attosecond physics experiments. Fig.~\ref{Fig.1.}(a) shows strong reduction in the harmonic intensity due to the ion motion. This reduction can be attributed to the change in the density and the amplitude of the oscillation electron layer, arising significantly due to the HB effect and to a lesser extent by electron-ion collisions. Moreover, very high plasma density does inhibit the formation and oscillation of the electron layer, and consequently poses a limit on HHG. This limit can be estimated by balancing the electrostatic force acting on the electron layer inside the plasma with the ponderomotive force of the laser, \text{i.e.} $n_{e}X_{\text{am}}\approx a_{{es}}\cos\theta\,\partial_{x}\, a_{{es}}$, where $a_{{es}}=a_{s}\exp{(-x/\lambda_{s})}$, $\lambda_{s}\propto n^{-1/2}_{e}$ is the skin-depth. This gives $X_{\text{am}}\approx 8I \text{cos}\,\theta\,/n^{3/2}_{e}$, showing that the amplitude  and consequently the harmonic intensity ($I_n$) vanishes at very high plasma density. Moreover, harmonic intensity ($I_n$) also decays with the harmonic frequency ($\omega_n$). Numerical calculations (not shown here) suggest the oscillating amplitude to be $X^{\text{min}}_{\text{am}} \ge 0.05\lambda_{l}$ in order to prevent the harmonic intensity from decaying faster than $I_n\propto \omega^{-8/3}_{n}$ \footnote{This calculation is based on the ROM model approximating the electron layer motion as a simple harmonic oscillator}. On equating $X_{\text{am}}$ with $X^{\text{min}}_{\text{am}}$, one can place an upper limit on the plasma density for intense HHG.   Consequently, with the given laser intensity $I$, the maximum density $n^{\text{max}}_{e}$ of the target plasma can be cast as
\begin{equation}
n_{e}^{\text{max}}=8.656I^{2/3}(\text{cos}\,\theta)^{-2/3}.
\label{eq3}
\end{equation}

\begin{figure}
  \includegraphics[width=0.45\textwidth,height=0.32\textwidth]{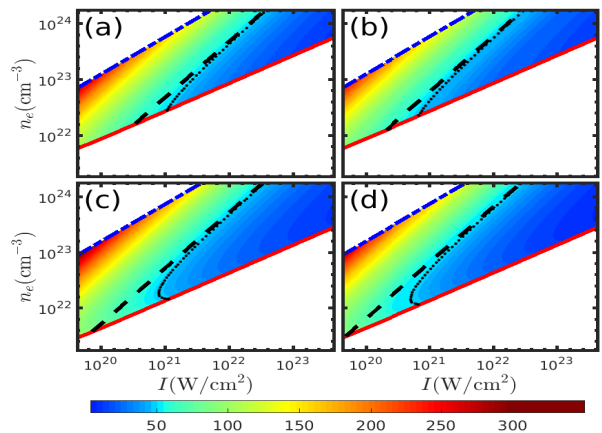}
  \caption{(color online) Parameter maps for optimum HHG for (a), (b) normal incidence and (c), (d) oblique incidence ($\theta=45^{\circ}$) with $p$-polarizations. (a), (c) are for gold plasma; (b), (d) are for carbon plasma. The color bar represents the hormonic numbers. The middle dotted and dashed black lines  are plotted for the $50$th harmonic in each case. See text for explanation.}\label{Fig.4.}
\end{figure}

With the help of the bounds given by relativistic critical density $n_{rc}$, Eqs.\eqref{eq2} and \eqref{eq3}, one can construct a parameter map relating the plasma density with the laser intensity and demarcate the map into two regions corresponding to both the coherent XUV radiation and an intense single attosecond pulse generations. Fig.~\ref{Fig.4.} shows these parameter maps for normal and oblique $(\theta=45^{\circ})$ incidences of the laser pulse on gold and carbon plasma targets. The lowest red line corresponds to $n_{rc}$ and sets a lower bound for HHG. The upper most dash-dotted blue line comes from Eq.\eqref{eq3} and it denotes the maximum density for intense HHG. The region between these two lines shows the contours of plasma densities given by Eq.\eqref{eq2} with color bar representing different harmonic orders. For instance, the middle dashed and dotted black lines have been plotted for the $50$th harmonic corresponding to a photon of energy $77.5$ eV in the XUV region of the electromagnetic spectrum. The middle dotted black line accounts for the hot electron generation and it deviates from the middle dashed black line only at lower plasma densities where the generation of the hot electron reduces the HB velocity significantly. Thus, the region of the map below the top most dash-dotted blue line, and above the intersection of middle dotted black and lower most red lines, is the region where overlapping between harmonic orders $(\le 50)$ is not significant, and it is suitable for generating coherent XUV radiations in the water-window region with high temporal coherence. While the region of the map between the middle dotted black and lower most red lines depicts a region where harmonic orders $(> 50)$ overlap significantly with each other, yielding a quasi-continuous spectra suitable for a single attosecond pulse generation.

To conclude, we have studied HHG in the ultra-relativistic regime of laser-plasma interaction. In this regime the ion motion, induced by the HB effect, tends to strongly affect the frequency bandwidth of the generated harmonics. The classical RR force does not strongly affect the frequency bandwidth but can lead to the reduction in the intensity of harmonics. Based on these considerations, we have scanned parameter maps (plasma density vs laser intensity) for different target materials at normal as well as oblique incidences of the laser pulse. These maps highlight the optimum regions for the generations of the coherent XUV radiation in the water-window part of the electromagnetic spectrum as well as an intense single attosecond pulse. These results are important for studies aimed at designing the next generation of short-wavelength radiation sources by employing the plasma high-harmonics.

One of the authors (NK) thanks Dr. Christian Ott for a useful discussion.


%

\end{document}